\documentclass[twocolumn,pra,showpacs,superscriptaddress]{revtex4}
\usepackage{amsmath}
\usepackage{graphicx}

\newcommand{\eq}{\begin{equation}}
\newcommand{\fine}{\end{equation}}

\begin{document}

\title{Anomalous resilient to decoherence macroscopic quantum superpositions  generated by universally covariant optimal quantum cloning}
\author{Nicol\`{o} Spagnolo}
\affiliation{Dipartimento di Fisica, ``Sapienza'' Universit\`{a} di Roma and Consorzio
Nazionale Interuniversitario per le Scienze Fisiche della Materia, Roma,
piazzale Aldo Moro 5, I-00185 Italy}
\author{Fabio Sciarrino}
\affiliation{Dipartimento di Fisica, ``Sapienza'' Universit\`{a} di Roma and Consorzio
Nazionale Interuniversitario per le Scienze Fisiche della Materia, Roma,
piazzale Aldo Moro 5, I-00185 Italy}
\affiliation{Istituto Nazionale di Ottica Applicata, largo Fermi 6, I-50125 Firenze, Italy}
\author{Francesco De Martini}
\affiliation{Dipartimento di Fisica, ``Sapienza'' Universit\`{a} di Roma and Consorzio
Nazionale Interuniversitario per le Scienze Fisiche della Materia, Roma,
piazzale Aldo Moro 5, I-00185 Italy}
\affiliation{Accademia Nazionale dei Lincei, via della Lungara 10, I-00165 Roma, Italy}

\begin{abstract}
We show that the quantum states generated by universal optimal quantum cloning of
a single photon represent an universal set of quantum superpositions resilient
to decoherence. We adopt Bures distance as a tool to investigate the persistence of
quantum coherence of these quantum states. According to this analysis,
the process of universal cloning realizes a class of quantum superpositions that exhibits
a covariance property in lossy configuration over the complete set of polarization
states in the Bloch sphere.
\end{abstract}

\maketitle

\section{Introduction}

In recent years two fundamental aspects of quantum mechanics have
attracted a great deal of interest, namely the investigation on the
irreducible nonlocal properties of Nature implied by quantum entanglement
and the physical realization of the \textquotedblleft Schr\oe dinger Cat
paradox\textquotedblright . The last concept, by applying the nonlocality
property to a combination of a microscopic and of a macroscopic systems,
enlightens the concept of the quantum state, the dynamics of large systems
and ventures into the most intriguing philosophical problem, i.e. the
emergence of quantum mechanics in the real life \cite{Schr35}. Unfortunately
it was always found extremely difficult to realize a system which realizes
simultaneously the following properties of the Schr\oe dinger Cat, i.e. a
Micro - Macroscopic Quantum Superposition (MMQS): (a) the quantum
superposition of two multiparticle, mutually orthogonal states, call it the
\textquotedblleft Macro-system\textquotedblright\ (b) the entanglement of
this superposition with a far apart single-particle state, i.e. the
\textquotedblleft Micro-system\textquotedblright . In addition, it was
always found that the quantum properties of any realized MMQS scheme were
quickly spoiled\ by the pervasive interactions with the environment: i.e. by
the effect of \textquotedblleft decoherence", the phase-disrupting effect
that so far has impaired the realization of the "quantum computer" \cite%
{Zure03}. The last crucial drawback made so far still more paradoxical any
MMQS scheme. Recently a MMQS realizing the conditions a) and b), consisting
of $\mathbf{N}\approx 3.5\times 10^{4}$ photons in a quantum superposition
and entangled with a far apart single - photon state was generated \cite%
{DeMa08}. The structure of this system was realized by means of the
quantum-injected optical parametric amplification (QI-OPA), i.e. a optimal
quantum-cloning machine. In addition and most surprisingly, our QI-OPA
system exhibited an anomalous large resilience to decoherence. 

In the present paper we demonstrate that the QI-OPA based MMQS is 
indeed a \textquotedblleft decoherence - free" system which, in particular,
is totally insensitive to temperature effects. This makes the device an
ideal approach to enlighten the quantum-to-classical transition and to
investigate the persistence of quantum phenomena into the \textquotedblleft
classical\textquotedblright\ domain by measurement procedures applied to
quantum systems of increasing size \cite{Zure03}. Furthermore, since the
generated Micro-Macro state is directly accessible at the output of the
apparatus, the implementation of significant multi qubit logic gates for
quantum information technology can be achieved by this method.

About ten years ago it was proposed to exploit the process of quantum
cloning to generate a different class of multiphoton states: Fig.1\cite%
{DeMa98,DeMa98a}. This method led recently to the successful experimental
realization of an entangled macroscopic quantum superposition (MQS) of a
large number of particles $N\approx 5\times 10^{4}$ \cite{Naga07,DeMa08}.
The persistence of quantum coherence in MQS states realized by the
\textquotedblleft phase-covariant" cloning, i.e. limited to a \textit{one
dimensional} subspace of the entire Bloch sphere of the Macro-qubit was
analyzed on the basis of two criteria based on the definition of "distance"
in the Hilbert space \cite{DeMa09,DeMa09a}. It was found that that limited
physical system shows a high resilience to decoherence at variance with any
coherent $|\alpha \rangle $ state MQS. The nice feature of phase-covariance
symmetry mostly consists of the relative simplicity of the required
"collinear" \ structure and of the high efficiency of the QI-OPA. This one
amplifies equally well all the single-photon polarization states $|\phi
\rangle $ belonging to the equatorial plane of the Bloch sphere of the
injected micro-qubit \cite{DeMa98a,Naga07}.

Given that lucky circumstance, a question arose whether it exists a physical
systems that exhibits the property of \textit{decoherence - freedom} in a
larger Hilbert space or, better, in the \textit{full space} available to the
generated Macrostate. The answer is yes, as demonstrated in the present
paper. The \textquotedblleft \textit{universal quantum cloning machine}"
realized in its \textquotedblleft \textit{optimal}" MMQS mode non-degenerate
configuration indeed possesses the requested property: the \textit{%
decoherence - freedom} is realized in the \textit{full Hilbert space}
spanned by the output Macrostate \cite{DeMa98,DeMa05,DeMa05a,DeMa07}.
\begin{figure}[t]
\centering
\includegraphics[width=0.5\textwidth]{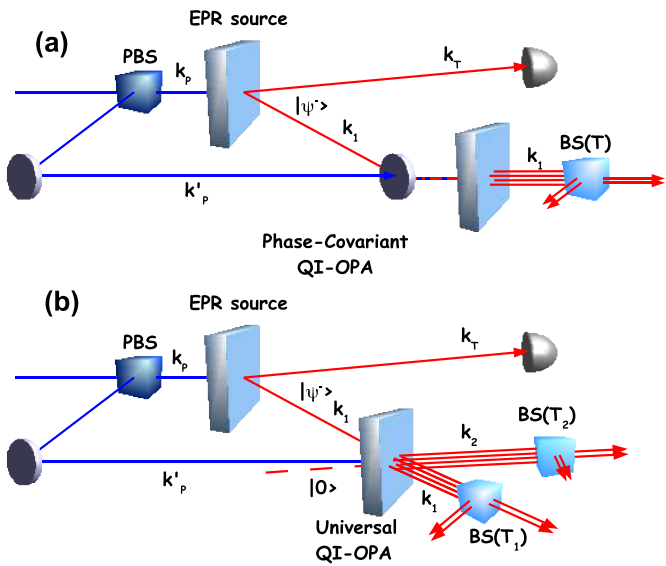}
\caption{(\textbf{a}) Scheme for the phase-covariant cloning of a single
photon state with a non-collinear optical parametric amplifier. The
beam-splitter [BS($T$)] is inserted to simulate the propagation over lossy
channels of the output field. (\textbf{b}) Scheme for the universal cloning
of a single photon state with a non-collinear optical parametric amplifier
(beam-splitters on spatial modes $\mathbf{k}_{1}$ [BS($T_{1}$)] and $\mathbf{%
k}_{2}$ [BS($T_{2}$)]). }
\label{fig:figure1}
\end{figure}

In this paper, we report the theoretical analysis on the resilience to
decoherence of the quantum states generated by universal quantum cloning of
a single-photon qubit. The basic tools of this investigation are provided by
the two coherence criteria defined in Refs. \cite{DeMa09,DeMa09a}. There,
the Bures distance \cite{Bure69,Hubn92,Jozs94}: 
\begin{equation}
\mathcal{D}\left( \hat{\rho},\hat{\sigma}\right) =\sqrt{1-\sqrt{\mathcal{F}(%
\hat{\rho},\hat{\sigma})}},
\end{equation}%
where $\mathcal{F}$ is a quantum ``fidelity", has been adopted as a measure
to quantify: (\textbf{I}) the ``\emph{distinguishability"} between two
quantum states $\left\{ |\phi _{1}\rangle ,|\phi _{2}\rangle \right\} $ and (%
\textbf{II)} the \emph{degree of coherence}, i.e. MQS visibility, of their
quantum superpositions $|\phi ^{+}\rangle =2^{-1/2}\left( |\phi _{1}\rangle
\pm |\phi _{2}\rangle \right) $. These criteria were chosen according to the
following considerations. \textbf{(I)} The distinguishability i.e. the
degree of orthogonality, represents the maximum discrimination power among
two quantum states available within a measurement. \textbf{(II)} The related 
$"Visibility"$ between the superpositions $|\phi ^{+}\rangle $ and $|\phi
^{-}\rangle $ depends exclusively on the relative phase of the component
states:$\ |\phi _{1}\rangle $ and $|\phi _{2}\rangle $. Consider two
orthogonal superpositions $|\phi ^{\pm }\rangle $: $\mathcal{D}\left( |\phi
^{+}\rangle ,|\phi ^{-}\rangle \right) =1$. In presence of decoherence the
relative phase between $\ |\phi _{1}\rangle $ and $|\phi _{2}\rangle $
progressively randomizes and the superpositions $|\phi ^{+}\rangle $ and $%
|\phi ^{-}\rangle $ approach an identical fully mixed state leading to: $%
\mathcal{D}\left( |\phi ^{+}\rangle ,|\phi ^{-}\rangle \right) =0$. The
physical interpretation of $\mathcal{D}\left( |\phi ^{+}\rangle ,|\phi
^{-}\rangle \right) $ as \textquotedblleft \textit{Visibility}%
\textquotedblright\ of a superposition $|\phi ^{\pm }\rangle $ is legitimate
insofar as the component states of the corresponding superposition, $|\phi
_{1}\rangle $ and $|\phi _{2}\rangle $ may be defined, at least
approximately, as \textquotedblleft \textit{pointer states}%
\textquotedblright\ or \textquotedblleft \textit{einselected states}%
\textquotedblright\ \cite{Zure03}. Within the set of the eigenstates
characterizing any quantum system the pointer states are defined as the ones
least affected by the external noise and that are highly resilient to
decoherence. In other words, the pointer states are \textquotedblleft quasi
classical\textquotedblright\ states which realize the minimum flow of
information from (or to) the System to (or from) the Environment. They are
invoved in all criteria of classicality, such as the ones based on
\textquotedblleft purity\textquotedblright\ and \textquotedblleft
predictability\textquotedblright\ of the macrostates \cite{Zure03}.

Our interest is aimed at the resilience properties of the different classes
of quantum states after the propagation over a lossy channel. This one is
modelled by a linear beam-splitter (BS) with transmittivity $T$ and
reflectivity $R=1-T$ acting on a state $\widehat{\rho}$ associated with a
single BS input mode: Fig.1. Let us first analyze the behaviour of the
coherent states and their superpositions. The investigation on the Glauber's
states leading to the $\alpha -MQS^{\prime }s$ case \cite{Schl91}:$|\Phi
_{\alpha \pm }\rangle $=$\mathcal{N}^{-1/2}\left( |\alpha \rangle \pm
|-\alpha \rangle \right) $ in terms of the \textquotedblleft pointer states" 
$|\pm \alpha \rangle \ $leads to the closed form result:\ $\mathcal{D}(|\Phi
_{\alpha +}\rangle ,|\Phi _{\alpha -}\rangle )=\sqrt{1-\sqrt{1-e^{-4R|\alpha
|^{2}}}}$ . This one is plotted in Fig.2 (dashed line) as function of \ the
average number of lost photons: $x\equiv R\langle n\rangle $. Note that the
value of $\mathcal{D}(|\Phi _{\alpha +}\rangle ,|\Phi _{\alpha -}\rangle )$
drops from 1 to 0.095 upon loss of only one photon: $x=1$. In other words,
the superpositions of $\alpha -states.$ $|\Phi _{\alpha \pm }\rangle $=$%
\mathcal{N}^{-1/2}\left( |\alpha \rangle \pm |-\alpha \rangle \right) $\
exhibit a fast decrease in their coherence, i.e. of their "visibility" and
"distinguishability", while the related components $|\pm \alpha \rangle $,
i.e. the \textquotedblleft pointer states\textquotedblright\ \cite{Zure03},
remain distinguishable until all photons of the state are depleted by the BS.



The amplification QI-OPA systems under investigation are reported in Fig.\ref%
{fig:figure1}. An EPR pair $|\psi ^{-}\rangle =2^{-1/2}\left( |H\rangle
_{1}|V\rangle _{T}-|V\rangle _{1}|H\rangle _{T}\right) $ is generated in a
first non-linear crystal. The symbols H and V refer to horizontal and
vertical field polarizations, i.e. the extreme \textquotedblleft
poles\textquotedblright\ of the Poincar\'{e} sphere. By analyzing and
measuring the polarization of the photon associated with the mode $\mathbf{k}%
_{T}$, the photon on mode $\mathbf{k}_{1}$ is nonlocally prepared in the
polarization qubit: $|\psi \rangle _{1}=\cos (\theta /2)|H\rangle
_{1}+e^{\imath \phi }\sin (\theta /2)|V\rangle _{1}$. Then, by an accurate
spatial focusing the single photon in the state $|\psi \rangle _{1}$ is
injected in the amplifier simultaneously with the strong UV pump beam $%
\mathbf{k^{\prime }}_{P}$. Let us analyze the two configurations of the
apparatus leading, as said, to two different regimes of quantum cloning:\
the "phase covariant" and the "universal".

\section{Phase-covariant optimal quantum cloning machine}

Let us first briefly review the results obtained for a "collinear" optical
configuration, i.e. leading to the phase-covariant optimal quantum cloning
machine: Figure 1 (a). The interaction Hamiltonian of this process is: $%
\widehat{\mathcal{H}}_{PC}=\imath \hbar \chi \widehat{a}_{H}^{\dag }\widehat{%
a}_{V}^{\dag }+\mathrm{H.c.}$ in the $\left\{ \vec{\pi}_{H},\vec{\pi}%
_{V}\right\} $ polarization basis, and $\widehat{\mathcal{H}}_{PC}=\frac{%
\imath \hbar \chi }{2}e^{-\imath \phi }\left( \widehat{a}_{\phi }^{\dag
\,2}-e^{\imath 2\phi }\widehat{a}_{\phi _{\bot }}^{\dag \,2}\right) +\mathrm{%
H.c.}$ for any \textquotedblright equatiorial\textquotedblright\ basis $%
\left\{ \vec{\pi}_{\phi },\vec{\pi}_{\phi \perp }\right\} $ on the Poincar%
\'{e} sphere having as "poles" the states: $\vec{\pi}_{H},\vec{\pi}_{V}$ .
The relevant equatorial basis considered here are $\left\{ \vec{\pi}_{+},%
\vec{\pi}_{-}\right\} $ and $\left\{ \vec{\pi}_{R},\vec{\pi}_{L}\right\} $
corresponding respectively to $\phi =0$ and $\phi =\pi /2$. The amplified
state for an injected equatorial qubit is: 
\begin{equation}
|\Phi _{PC}^{\phi }\rangle =\sum_{i,j=0}^{\infty }\gamma _{ij}|(2i+1)\phi
,(2j)\phi _{\bot }\rangle  \label{eq:Phi_equat}
\end{equation}%
where $\gamma _{ij}=\frac{1}{C^{2}}\left( e^{-\imath \varphi }\frac{\Gamma }{%
2}\right) ^{i}\left( -e^{\imath \varphi }\frac{\Gamma }{2}\right) ^{j}\frac{%
\sqrt{(2i+1)!}\,\sqrt{(2j)!}}{i!j!}$, $C=\cosh g$ , $\Gamma =\tanh g$.
Hereafter, $|p\psi ,q\psi _{\bot }\rangle _{i}$ stands for a Fock state with 
$p$ photons polarized $\vec{\pi}_{\psi }$ and $q$ photons polarized $\vec{\pi%
}_{\psi _{\bot }}$ on spatial mode $\mathbf{k}_{i}$. We evaluated
numerically the \textit{distinguishability} of $\left\{ |\Phi
_{PC}^{+,-}\rangle \right\} $ through the distance $\mathcal{D}(|\Phi
_{PC}^{+}\rangle ,|\Phi _{PC}^{-}\rangle )$: Fig.2. Consider the MQS of the
macrostates $|\Phi _{PC}^{+}\rangle $, $|\Phi _{PC}^{-}\rangle $: $|\Psi
_{PC}^{\pm }\rangle =\frac{\mathcal{N}_{\pm }}{\sqrt{2}}\left( |\Phi
_{PC}^{+}\rangle \pm i|\Phi _{PC}^{-}\rangle \right) $. Due to the linearity
of theamplification process \cite{DeMa05}, $|\Psi _{PC}^{\pm }\rangle
=\left\vert \Phi _{PC}^{R/L}\right\rangle $ and in virtue of the
phase-covariance of the process:

\begin{equation}
\mathcal{D}(|\Psi _{PC}^{+}\rangle ,|\Psi _{PC}^{-}\rangle )=\mathcal{D}%
(|\Phi _{PC}^{R}\rangle ,|\Phi _{PC}^{L}\rangle )=\mathcal{D}(|\Phi
_{PC}^{+}\rangle ,|\Phi _{PC}^{-}\rangle )
\end{equation}
These equations can be assumed as the theoretical conditions assuring the 
\textit{decoherence - freedom} of any quantum MMQS\ state generated by the
QI-OPA\ in the collinear configuration: they identify the "equatorial set"
of the Bloch sphere a privileged \textit{decoherence-free} Hilbert subspace.
The \textit{visibility} of the q-MQS $\left\{ |\Psi _{PC}^{+,-}\rangle
\right\} $ was evaluated numerically analyzing the Bures distance as a
function of $x$. Note that for small values of $x$ the decay of $D(x)$ is
far slower than for the coherent $\alpha -MQS$ case. This resilience to
decoherence feautre has been experimentally verified in \cite{DeMa08}.%

\begin{figure}[t!!]
\centering
\includegraphics[width=0.35\textwidth]{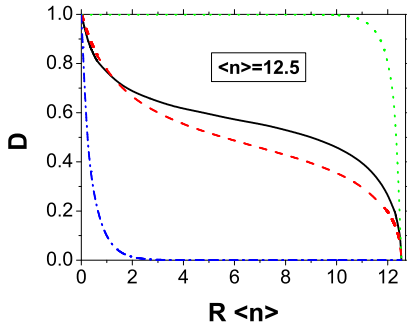}
\caption{Bures distance for various classes of MQSs for $\langle n\rangle
=12.5$. The lower blue dash-dotted curve corresponds to the quantum
superpositions of coherent states $|\protect\alpha \rangle \pm |-\protect%
\alpha \rangle $, while the green dotted upper curve is relative to the
distinguishability between the component states $|\pm \protect\alpha \rangle 
$. The black straight curve corresponds to the MQS generated by
phase-covariant cloning \protect\cite{DeMa09,DeMa09a}, while the red dashed
curve corresponds to the universal cloning based MQS.}
\label{fig:figure5}
\end{figure}

\section{Universal optimal quantum cloning machine}

Let's now investigate the resilience to decoherence of the MMQS generated by
the \textit{universal optimal quantum cloning machine} (UOQCM). According to
the original proposal, this was implemented experimentally by QI-OPA\ device
working in a non - collinear, i.e. mode non-degenerate configuration:\
Figure 2 (b) \cite{DeMa98}. 
This parametric amplifier acts as an universal $N\rightarrow M$ quantum
cloning machine \cite{Pell03,DeMa04} as well as a \textit{Universal - Not}
(U-Not) quantum machine \cite{DeMa02}. The interaction Hamiltonian for the
amplifier is now given by $\hat{\mathcal{H}}_{U}=\imath \hbar \chi (\hat{a}%
_{1\psi }^{\dag }\hat{a}_{2\psi _{\bot }}^{\dag }-\hat{a}_{1\psi _{\bot
}}^{\dag }\hat{a}_{2\psi }^{\dag })+\mathrm{H.c.}$, where $\vec{\pi}_{\psi
}=\cos (\theta /2)\vec{\pi}_{H}+e^{\imath \phi }\sin (\theta /2)\vec{\pi}%
_{V} $ and $\vec{\pi}_{\psi _{\bot }}=(\vec{\pi}_{\psi })^{\bot }$. Since
this system possesses a complete SU(2) simmetry, the Hamiltonian maintains
the same form for any simultaneous rotation of the Bloch sphere of the
polarization basis for both output modes $\mathbf{k}_{1}$ and $\mathbf{k}%
_{2} $. 

\begin{widetext}

\begin{figure}[t!!]
	\centering
		\includegraphics[width=0.90\textwidth]{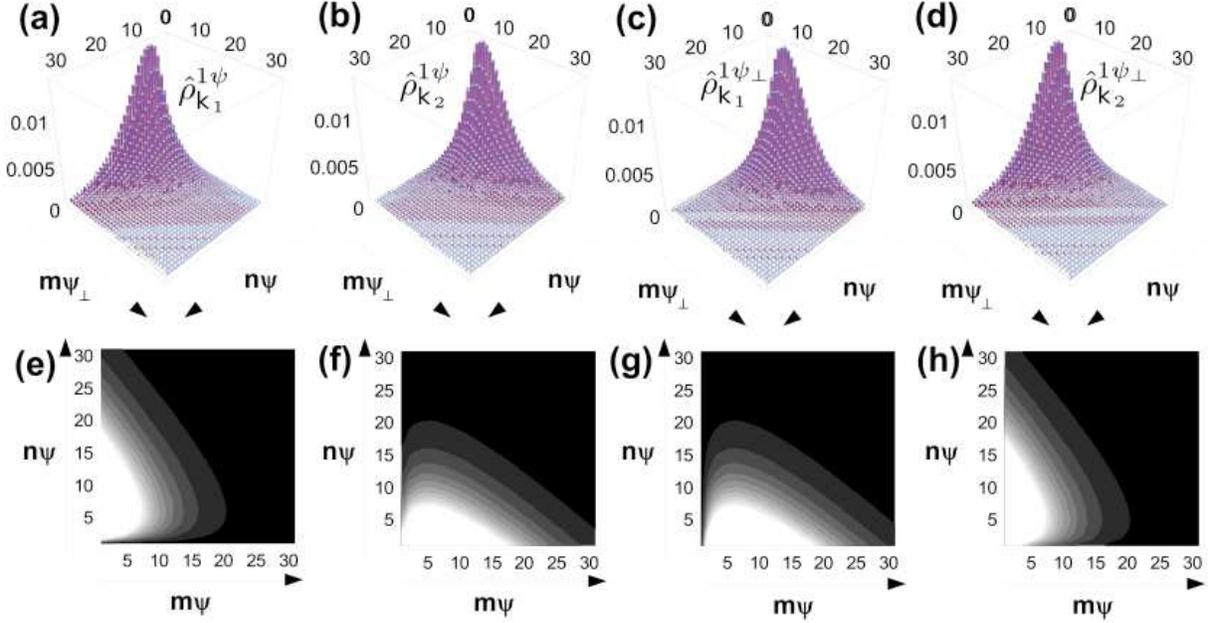}
	\caption{Probability distribution (a-d) and contour plots (e-h) of the reduced 
	density matrices $\hat{\rho}_{\mathbf{k}_{1}}^{1 \psi}$ (a)-(e), 
	$\hat{\rho}_{\mathbf{k}_{2}}^{1 \psi}$ (b)-(f), $\hat{\rho}_{\mathbf{k}_{1}}^{1 \psi_{\bot}}$ 
	(c)-(g) and $\hat{\rho}_{\mathbf{k}_{2}}^{1 \psi_{\bot}}$ (d)-(h). All plots correspond to the
	gain value $g=1.5$.}
	\label{fig:figure2}
\end{figure}

\end{widetext}

The output state of the amplifier reads: 
\begin{equation}
\begin{aligned} \vert \Phi^{1 \psi}_{U} \rangle &= \hat{U} \vert 1 \psi
\rangle_{1} = \frac{1}{C^{3}} \sum_{n,m=0}^{\infty} \Gamma^{n+m} (-1)^{m}
\sqrt{n+1} \\ &\vert (n+1)\psi, m\psi_{\bot} \rangle_{1} \otimes \vert
m\psi,n\psi_{\bot} \rangle_{2} \end{aligned}
\label{eq:amplified_state_no_losses}
\end{equation}%
In order to investigate the features of the state of Eq.(\ref%
{eq:amplified_state_no_losses}), Fig.\ref{fig:figure2} reports the
photon-number distribution for the reduced states $\hat{\rho}_{U\mathbf{k}%
_{i}}^{1\psi (1\psi _{\bot })}$= $\mathrm{Tr}_{\mathbf{k}_{i}}\left[ |\Phi
_{U}^{1\psi (1\psi _{\bot })}\rangle \langle \Phi _{U}^{1\psi (1\psi _{\bot
})}|\right] $ associated to the output spatial modes $\mathbf{k}_{1}$ and $%
\mathbf{k}_{2}$. The photon-number distributions in the $\mathbf{k}_{1}$
spatial mode [Figs.\ref{fig:figure2}-(a) and (c)], i.e. the cloning mode,
show a strong unbalancement along the direction of the injected polarization
state. The anticloning $\mathbf{k}_{2}$ mode [Figs.\ref{fig:figure2}-(b) and
(d)] presents the opposite unbalancement along the direction of the
orthogonal polarization, since on that spatial mode the amplifier works as a 
\textit{U-Not machine} \cite{DeMa02}. This feature is also enlightened by
the contour plots of Figs.\ref{fig:figure2}-(e-h), where the white regions
represent the Fock-space zones where the photon-number distributions are
more densely populated. Furthermore, we note that at variance with the
phase-covariant amplifier \cite{DeMa09,DeMa09a}, the output states do not
exhibit any comb structure in their photon number distributions. In
agreement with the "universality" property of the source, the Bures distance
between the MQS states $|\Phi _{U}^{1\psi }\rangle =\cos (\theta /2)|\Phi
_{U}^{1H}\rangle +e^{\imath \phi }\sin (\theta /2)|\Phi _{U}^{1V}\rangle $
and $|\Phi _{U}^{1\psi _{\bot }}\rangle $ is independent on the choice of $%
(\theta ,\phi )$:

\begin{equation}
\mathcal{D}(\hat{\rho}_{U}^{1\psi },\hat{\rho}_{U}^{1\psi _{\bot }})=%
\mathcal{D}(\hat{\rho}_{U}^{1\psi ^{\prime }},\hat{\rho}_{U}^{1\psi _{\bot
}^{\prime }})
\end{equation}

for any basis $\left\{ \vec{\pi}_{\psi },\vec{\pi}_{\psi ^{\prime }}\right\} 
$. This feature is the extension of the $\phi $-covariance property of the
collinear quantum cloning machine \cite{DeMa09a} to the full set of
polarization states on the output Bloch spheres..

\begin{figure}[t!]
\centering
\includegraphics[width=0.5\textwidth]{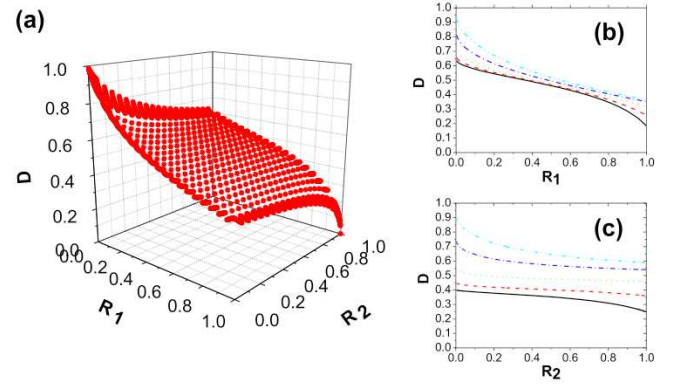}
\caption{Bures distance between the MQSs $\vert \Phi^{1 \protect\psi}
\rangle $ and $\vert \Phi^{1 \protect\psi_{\bot}} \rangle$ after losses. (a)
3-dimensional surface $\mathcal{D}(R_{1},R_{2})$ as a function of the two
spatial mode parameters $R_{1} = 1-T_{1}$ and $R_{2}=1-T_{2}$. (b) $\mathcal{%
D}(R_{1},R_{2})$ with $R_{2}$ fixed, as a function of $R_{1}$. (c) $\mathcal{%
D}(R_{1},R_{2})$ with $R_{1}$ fixed, as a function of $R_{2}$. (b)-(c)
Straight curves correspond to $R_{2(1)}=0.9$, the dashed curves to $%
R_{2(1)}=0.75$, dotted one to $R_{2(1)}=0.5$, the dash-dotted curves to $%
R_{2(1)}=0.2$ and dash-dot-dotted curves to $R_{2(1)}=0.05$.}
\label{fig:figure3}
\end{figure}

By introducing, in analogy with the previous case, two beam-splitters ($%
BS_{1}$ and $BS_{2}$)\ on the output states, we evaluate by standard
algebraic numerical routines the Bures distance between the orthogonal
macrostates $|\Phi _{U}^{1\psi }\rangle $ and $|\Phi _{U}^{1\psi _{\bot
}}\rangle $ as a functions of the corresponding transmission parameters: $%
T_{1}$ and $T_{2}$. In Fig.\ref{fig:figure3}-(a) we report the 3-dimensional
plot of the function $\mathcal{D}(R_{1},R_{2})=\mathcal{D}(\hat{\rho}^{1\psi
},\hat{\rho}^{1\psi _{\bot }})$ for a gain value of $g=1.2$, corresponding
to an overall average number of photons $\langle \hat{n}\rangle
=\sum_{i=1}^{2}\left[ \langle \hat{n}_{i\psi }\rangle +\langle \hat{n}%
_{i\psi _{\bot }}\rangle \right] \approx 15$. The Figure shows that the MQS
visibility possesses a resilient structure in presence of losses, since the
Bures distance does not decrease exponentially with the lossy parameters $%
\left\{ R_{1},R_{2}\right\} $. In Figs.\ref{fig:figure3}-(b-c) we then
report several sections of the 3-dimensional surface of Fig.\ref{fig:figure3}%
-(a) by fixing either $R_{1}$ or $R_{2}$. We note that the $|\Phi
_{U}^{1\psi }\rangle $ and $|\Phi _{U}^{1\psi _{\bot }}\rangle $ MQSs are
more sensitive to losses in the \textit{cloning} mode: $\mathbf{k}_{1}$ than
in the \textit{anticloning} one: $\mathbf{k}_{2}$. This can be explained by
considering that the distance between these orthogonal Mcro-states depends
on the unbalancement in the corresponding photon-number distributions. Since
this feature is pronounced in the spatial cloning mode $\mathbf{k}_{1}$,
losses acting on this mode cancel more rapidly the orthogonality between $%
|\Phi _{U}^{1\psi }\rangle $ and $|\Phi _{U}^{1\psi _{\bot }}\rangle $. As
the number of photons present in the state is increased, the visibility
keeps large up to a value$\ V\approx 0.5$ in a larger range of the number of
reflected photons. All this shows that, in analogy with the phase-covariant
case, the MQS's generated by quantum cloning become more resilient to losses
since the quantum coherence present in these state can survive the loss of a
larger number of photons.

\begin{figure}[tbp]
\centering
\includegraphics[width=0.35\textwidth]{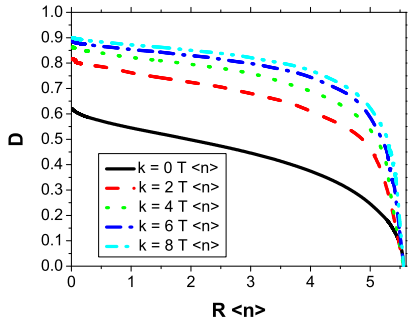}
\caption{Numerical results for the Bures distance between the $\vert \Phi^{1%
\protect\psi} \rangle$ and $\vert \Phi^{1 \protect\psi_{\bot}} \rangle$
states after the propagation over a lossy channel and the application of the
O-Filter device. From the lower to the upper curve, the filtering threshold
is set to $k = 0 \langle n \rangle$ ($P_{filt} = 1$), $k = 2 \langle n
\rangle$ ($P_{filt} = 1.7 \times 10^{-1}$), $k = 4 \langle n \rangle$ ($%
P_{filt} = 1.5 \times 10^{-2}$), $k = 6 \langle n \rangle$ ($P_{filt} = 1.1
\times 10^{-3}$), $k = 8 \langle n \rangle$ ($P_{filt} = 8.2 \times 10^{-5}$%
). All curves are calculated for a gain $g=1.2$, corresponding to an average
number of generated photons on mode $\mathbf{k}_{1}$ equal to $\langle n
\rangle \simeq 8$.}
\label{fig:figure6}
\end{figure}

Finally we analyze the action of the Orthogonality Filter (O-Filter, OF) on
the amplified single photon states $|\Phi _{U}^{1\psi }\rangle $ by
analyzing how the Bures distance is affected by the application of this
device. 
The POVM like technique \cite{Pere95} implied by this device locally selects
the events for which the difference between the photon numbers associated
with two orthogonal polarizations $|m-n|>k$, i.e. larger than an adjustable
threshold, $k$ \cite{Naga07}. By this method a sharper discrimination
between the output states $|\Phi _{U}^{1 \psi}\rangle $ and $|\Phi _{U}^{1 \psi
_{\bot }}\rangle $ can be achieved. 
\ We focus our attention only on the reduced density matrix $\hat{\rho}_{%
\mathbf{k}_{1}}^{1\psi }(T)$ corresponding to the output spatial mode $%
\mathbf{k}_{1}$. We then applied the numerical methods previously adopted in
order to calculate the Bures distance $\mathcal{D}(x)$, where $x=R\langle
n\rangle $ is the number of lost photons, as a function of the threshold $k$%
. In Fig.\ref{fig:figure6} the results of a numerical analysis carried out
for $g=1.2$ and different values of $k$\ are reported.

Note the increase of the value of $\mathcal{D}(x)$, i.e. of the MQS
Visibility, by increasing $k$. Of course, in the spirit of any POVM
measurement, the high interference visibility is here achieved at the cost
of a lower success probability \cite{Hutt96}. The general, most important
feature shown by all these Figures is that both the \textquotedblleft 
\textit{Distinguishability}\textquotedblright\ and the \textquotedblleft 
\textit{Visibility}\textquotedblright\ of all \textquotedblleft
universal\textquotedblright\ macrostates $|\Phi _{U}^{1\psi }\rangle ,|\Phi
_{U}^{1\psi \bot }\rangle $ as well as of all their \textquotedblleft
universal\textquotedblright\ quantum superpositions can be kept close to the
maximum value in spite of the increasing effect of decoherence due to
increasing values of the $BS_{1}$reflectivity: $R\langle n\rangle $. On the
basis of all these results we may then conclude that all the
\textquotedblleft universal\textquotedblright\ macrostates and
superpositions generated by the QI-OPA may be safely defined as classically
stable, \textit{einselected} \textquotedblleft \textit{pointer states}%
\textquotedblright\ \cite{Zure03}.

\section{Conclusions}

The present paper reports a thorough theoretical analysis on the 
resilience to decoherence of the quantum superpositions generated by
universal quantum cloning. This property is found to
depend on the symmetry (\textquotedblleft covariance") of the optimal
cloning process, which allows to identify a covariant set of stable quantum
superpositions over the full Hilbert space spanned by the generated
Macro-states. In order to gain insight into the general picture and to
support the congruence of our final conclusions we find useful to relate
here the various aspects of the optimal cloning process with the current
MQS\ physical model considered by \cite{Zure03}.

1) The ``\textit{System}'' in our optical entangled amplification scheme is
represented by the assembly of $N+1$ photon particles associated with the
macrostate $|\Phi ^{\phi }\rangle$ generated by the optimal cloning
apparatus.

2) The flow of (classical) ``noise information'' directed from the ``\textit{%
Environment}'' towards the \textit{System} is attributed to the unavoidable
squeezed-vacuum noise affecting the building up of the macrostate $|\Phi
^{\phi }\rangle $ within the process of amplification. \ As already
stressed, the ``\textit{optimality}'' of the quantum cloning generally
implies that the flow of classical noise is the \textit{minimum} allowed by
the principles of quantum mechanics, i.e. by the ``\textit{no-cloning theorem%
}'' \cite{Scia05,Scia07}.

3) The flow of quantum information directed from the System towards the
Environment is provided by the controlled \textquotedblleft decoherence in
action\textquotedblright\ provided by the BS-scattering process and by the
losses taking place in all photo-detection processes. We have seen that by
the use of the OF, or even in the absence of it, the interference
phase-distrupting effects caused by the decoherence can be efficiently tamed
for the macrostates and for their quantum superpositions.

4) By the universal cloning method considered here the maximum allowed 
\textit{Distinguishability} and \textit{Visibility }are attained\textit{\ }%
for the macro-qubit\ Hilbert space.

5) In any cloning apparatus a unitary transformation $\hat{U}$ connects all
physical properties belonging to the micro-world to the corresponding ones
belonging to the macrosopic \textquotedblleft classical\textquotedblright\
world. Any lack of perceiving this close correspondence, for instance in
connection with the realization of the \textquotedblleft Schr\"{o}dinger
Cat\textquotedblright\ must be only attributable to the intrinsic
limitations of our perceiving senses, of our observational methods or of our
measurement apparata. In other words, at least in our case, the two worlds
are deterministically mirrored one into the other by the unitary map $\hat{U}
$ \ which is provided by quantum mechanics itself. This is the key to
understand our results.

6) The q-MQS based on the cloning process is not a \textquotedblleft
thermodynamic\textquotedblright\ system as its dynamics and decoherence do
not depend on the temperature T. It rather belongs to, and indeed
establishes a first and most insightful physical model of, the class of the
\textquotedblleft \textit{parametrically - driven, open quantum -
statistical systems}\textquotedblright\ that have been recently invoked\ to
provide and sustain\ extended long - range nonlocal coherence processes in
complex biological photosyntetic systems \cite{Cai08,Enge07}

\begin{appendix}

\section{Calculation of the density matrix coefficients for the universally
amplified single-photon states in presence of losses}
\label{app:matrix}

In this appendix we report the detailed calculation of the coefficient of the density
matrix for the $\vert \Phi^{1 \psi} \rangle$ states in presence of losses.
We focus our attention on the $|1\psi \rangle _{1}$ case
only, since the calculation for the complementary state $|1\psi _{\bot
}\rangle _{1}$ is similar.

First we investigate the features of the interaction
Hamiltonian. Due to the properties of $\hat{\mathcal{H}}_{U}$, the time
evolution operator in the interaction picture $\hat{U}=\exp (-\imath \hat{%
\mathcal{H}}_{U}t/\hbar )$ can be decomposed as the product of two
independent operators $\hat{U}=\hat{U}_{\mathcal{A}}\otimes \hat{U}_{%
\mathcal{A^{\prime }}}$, acting on two different Hilbert spaces
corresponding to the two sets of modes $\mathcal{A}\equiv \{(\mathbf{k}_{1},%
\vec{\pi}_{\psi }),(\mathbf{k}_{2},\vec{\pi}_{\psi _{\bot }})\}$ and $%
\mathcal{A^{\prime }}\equiv \{(\mathbf{k}_{1},\vec{\pi}_{\psi _{\bot }}),(%
\mathbf{k}_{2},\vec{\pi}_{\psi })\}$ \cite{Pell03,DeMa04}: 
\begin{eqnarray}
\hat{U}_{\mathcal{A}} &=&\exp \left[ \chi t(\hat{a}_{1\psi }^{\dag }\hat{a}%
_{2\psi _{\bot }}^{\dag }-\hat{a}_{1\psi }\hat{a}_{2\psi _{\bot }})\right] \\
\hat{U}_{\mathcal{A^{\prime }}} &=&\exp \left[ -\chi t(\hat{a}_{1\psi _{\bot
}}^{\dag }\hat{a}_{2\psi }^{\dag }-\hat{a}_{1\psi _{\bot }}\hat{a}_{2\psi })%
\right]
\end{eqnarray}%
In the case of a separable input state in the QI-OPA $\hat{\rho}=\hat{\rho}_{%
\mathcal{A}}\otimes \hat{\rho}_{\mathcal{A^{\prime }}}$, the amplified
states can be written in a separable form: 
\begin{equation}
\hat{\rho}(t)=\hat{U}\hat{\rho}\hat{U}^{\dag }=\left( \hat{U}_{\mathcal{A}}%
\hat{\rho}_{\mathcal{A}}\hat{U}_{\mathcal{A}}^{\dag }\right) \otimes \left( 
\hat{U}_{\mathcal{A^{\prime }}}\hat{\rho}_{\mathcal{A^{\prime }}}\hat{U}_{%
\mathcal{A^{\prime }}}^{\dag }\right)  \label{eq:separability_property}
\end{equation}%
This separability property will be exploited in the remaining part of the
paper both for the calculation of the density matrix after losses and for
the evaluation of the Bures distance. 

\subsection{Density matrix coefficients for the amplified
states}

In this section we derive the density matrix coefficients for the amplified
states $|\Phi ^{1\psi (1\psi _{\bot })}\rangle $ after the transmission over
a lossy channel. We focus our attention on the $|1\psi \rangle _{1}$ case
only, since the calculation for the complementary state $|1\psi _{\bot
}\rangle _{1}$ is similar. Due to the property of the ``universal" amplifier
analyzed previously, we analyze separately the two
subspaces $\mathcal{A}$ and $\mathcal{A^{\prime }}$ [Eq.(\ref%
{eq:separability_property})]. Since the time evolution operators $\hat{U}_{%
\mathcal{A}}$ and $\hat{U}_{\mathcal{A^{\prime }}}$ are equivalent apart
from a global phase factor $(-1)$, the quantum states for the amplifier $%
\mathcal{A^{\prime }}$ can be directly derived from the expressions obtained
for amplifier $\mathcal{A}$. Only two relevant cases are considered: the
vacuum injected state $\hat{U}_{\mathcal{A}}|0\rangle $ (spontaneous
emission) and the single-photon injected $\hat{U}_{\mathcal{A}}|1\psi
\rangle _{1}$ state. This analysis can be performed separately for the two
amplifiers since the separability feature also holds after the amplified
state propagates over a lossy channel in both $\mathbf{k}_{1}$ and $\mathbf{k%
}_{2}$ spatial modes. This is a consequence of the properties of the lossy
channel map, which being a ``local" transformation, acts independently on
each mode. The output state after losses reads: 
\begin{equation}
\mathcal{L}[\hat{\rho}(t)]=\mathcal{L}_{\mathcal{A}}\left[ \hat{U}_{\mathcal{%
A}}\hat{\rho}_{\mathcal{A}}\hat{U}_{\mathcal{A}}^{\dag }\right] \otimes 
\mathcal{L}_{\mathcal{A^{\prime }}}\left[ \hat{U}_{\mathcal{A^{\prime }}}%
\hat{\rho}_{\mathcal{A^{\prime }}}\hat{U}_{\mathcal{A^{\prime }}}^{\dag }%
\right]  \label{eq:disentangling}
\end{equation}%
Here $\mathcal{L}_{\mathcal{A}}=\mathcal{L}_{\mathbf{k}_{1},\vec{\pi}_{\psi
}}\otimes \mathcal{L}_{\mathbf{k}_{2},\vec{\pi}_{\psi _{\bot }}}$, $\mathcal{%
L}_{\mathcal{A^{\prime }}}=\mathcal{L}_{\mathbf{k}_{1},\vec{\pi}_{\psi
_{\bot }}}\otimes \mathcal{L}_{\mathbf{k}_{2},\vec{\pi}_{\psi }}$ are the
maps induced by losses for the two subspaces, where the single mode map 
$(\mathbf{k}_{i},\vec{\pi})$ is given by the following expression \cite{Durk04}: 
\begin{equation}
\label{eq:lossy_channel_map}
\mathcal{L}_{\mathbf{k}_{i},\vec{\pi}}[\hat{\sigma}]=\sum_{p=0}^{\infty
}R_{i}^{p/2}T_{i}^{(\hat{a}^{\dag }\hat{a})/2}\frac{\hat{a}_{\mathbf{k}_{i},%
\vec{\pi}}^{p}}{\sqrt{p!}}\hat{\sigma}\frac{\hat{a}_{\mathbf{k}_{i},\vec{\pi}%
}^{\dag \,p}}{\sqrt{p!}}T_{i}^{(\hat{a}^{\dag }\hat{a})/2}R_{i}^{p/2}
\end{equation}%
with $T_{i}$ the transmission efficiency of the channel, assumed to be
polarization independent.

We begin with the analysis of the spontaneous emission regime. The calculation
proceeds as follows. Starting from the quantum state for the subsystem $\mathcal{A}$ 
$\hat{U}_{\mathcal{A}} \vert 0 \rangle$, the output state after the transmission over 
the lossy channel is obtained by applying the lossy channel map (\ref{eq:lossy_channel_map}) 
to the density matrix of the state $\hat{\rho}_{\mathcal{A}}^{0}$. The same procedure 
applies for the single photon amplified states,
where the seed of the amplifier $\mathcal{A}$ is the single photon state $%
\vert 1 \psi \rangle$. In this case, the input state in the lossy channel
is $\hat{U}_{\mathcal{A}}$. By applying the lossy channel map over the density matrix $\hat{\rho}_{%
\mathcal{A}}^{1 \psi}$ of the state, we find the desired output states.
Details on the calculation and the complete expressions of the coefficients for
the density matrices $\hat{\rho}^{1\psi}_{\mathcal{A}}(T_{1},T_{2})$ and 
$\hat{\rho}_{\mathcal{A}}^{0}(T_{1},T_{2})$ are reported below.

Let us emphasize that, due to analogy of the Hamiltonian of the two
amplifier $\mathcal{A}$ and $\mathcal{A^{\prime}}$, the density matrices of
the states $\hat{\rho}_{\mathcal{A^{\prime}}}^{0} (T_{1},T_{2})$ and $\hat{%
\rho}_{\mathcal{A^{\prime}}}^{1 \psi_{\bot}} (T_{1},T_{2})$ for amplifier $%
\mathcal{A^{\prime}}$ can be directly derived from Eqs.(\ref%
{eq:losses_spontaneous_1}-\ref{eq:losses_spontaneous_3}) and (\ref%
{eq:losses_amplified_1}-\ref{eq:losses_amplified_3}) by substituting $(
\Gamma)$ with $(- \Gamma)$ and by re-labelling the indexes describing the
spatial and polarization modes.

We begin with the analysis of the spontaneous emission regime. The quantum
state for the subsystem $\mathcal{A}$ is given by: 
\begin{equation}
\hat{U}_{\mathcal{A}} \vert 0 \rangle = \frac{1}{C} \sum_{n=0}^{\infty}
\Gamma^{n} \vert n\psi \rangle_{1} \otimes \vert m \psi_{\bot} \rangle_{2}
\end{equation}
The output state after the transmission over the lossy channel is obtained
by applying the lossy channel map (\ref{eq:lossy_channel_map}) to the
density matrix of the state $\hat{\rho}_{\mathcal{A}}^{0}$: 
\begin{equation}
\hat{\rho}_{\mathcal{A}}^{0} (T_{1}, T_{2})= \left( \mathcal{L}_{\mathbf{k}%
_{1},\vec{\pi}_{\psi}} \otimes \mathcal{L}_{\mathbf{k}_{2}, \vec{\pi}%
_{\psi_{\bot}}} \right) \left[ \hat{\rho}_{\mathcal{A}}^{0} \right]
\end{equation}
After direct application of the lossy channel map on the density matrix, the
following expression is obtained: 
\begin{widetext}
\begin{equation}
\label{eq:losses_spontaneous_1}
\begin{aligned}
\hat{\rho}_{\mathcal{A}}^{0}(T_{1}, T_{2}) &= \sum_{i=0}^{\infty} \sum_{j=0}^{i}
\sum_{k=i-j}^{\infty} \big[ \hat{\rho}_{\mathcal{A}}^{0}(T_{1}, T_{2}) 
\big]_{ijk}^{(i \geq j)} \vert i \psi \rangle_{1} \langle k \psi \vert \otimes
\vert j \psi_{\bot} \rangle_{2} \langle (j+k-i)\psi_{\bot}\vert +\\
&+ \sum_{i=0}^{\infty} \sum_{j=i+1}^{\infty} \sum_{k=0}^{\infty} \big[ 
\hat{\rho}_{\mathcal{A}}^{0}(T_{1}, T_{2}) \big]_{ijk}^{(i < j)} 
\vert i \psi \rangle_{1} \langle k \psi \vert \otimes \vert j \psi_{\bot} 
\rangle_{2} \langle (j+k-i)\psi_{\bot}\vert
\end{aligned}
\end{equation}
where the coefficients for $i\geq j$ and $i<j$ are given by:
\begin{eqnarray}
\big[ \hat{\rho}_{\mathcal{A}}^{0}(T_{1}, T_{2}) 
\big]_{ijk}^{(i \geq j)} &=& \frac{1}{C^{2}} \Gamma^{i+k}  \frac{T_{1}^{(i+k)/2}
T_{2}^{(2j+k-i)/2} R_{2}^{i-j} \sqrt{i!k!}}{(i-j)! \sqrt{j!(j+k-i)!}} 
\, _{2}F_{1}\left(1+i,1+k,i+i-j;\Gamma^{2} R_{1} R_{2} \right)\\
\label{eq:losses_spontaneous_3}
\big[ \hat{\rho}_{\mathcal{A}}^{0}(T_{1}, T_{2}) 
\big]_{ijk}^{(i < j)} &=& \frac{1}{C^{2}} \Gamma^{i+k}  \frac{T_{1}^{(i+k)/2}
T_{2}^{(2j+k-i)/2} R_{1}^{j-i} \sqrt{j!(j+k-i)!}}{(j-i)! \sqrt{i!k!}} 
\, _{2}F_{1}\left(1+j,1+j+k-i,1+j-i;\Gamma^{2} R_{1} R_{2} \right)
\end{eqnarray}
\end{widetext}
where $_{2}F_{1}(a,b,c;z)$ is the hypergeometric function defined in Ref. 
\cite{Slat66}. The same procedure has been applied to the stimulated case,
where the seed of the amplifier $\mathcal{A}$ is the single photon state $%
\vert 1 \psi \rangle$. In this case, the input state in the lossy channel
has the following expression: 
\begin{equation}
\hat{U}_{\mathcal{A}} \vert 1 \psi \rangle_{1} = \frac{1}{C^{2}}
\sum_{n=0}^{\infty} \Gamma^{n} \sqrt{n+1} \vert (n+1)\psi \rangle_{1}
\otimes \vert m \psi_{\bot} \rangle_{2}
\end{equation}
By applying the lossy channel map over the density matrix $\hat{\rho}_{%
\mathcal{A}}^{1 \psi}$ of the state, we find: 
\begin{equation}
\hat{\rho}_{\mathcal{A}}^{1 \psi} (T_{1}, T_{2})= \left( \mathcal{L}_{%
\mathbf{k}_{1},\vec{\pi}_{\psi}} \otimes \mathcal{L}_{\mathbf{k}_{2}, \vec{%
\pi}_{\psi_{\bot}}} \right) \left[ \hat{\rho}_{\mathcal{A}}^{1 \psi} \right]
\end{equation}
The application of the map leads to the following expression for the density
matrix: 
\begin{widetext}
\begin{equation}
\begin{aligned}
\label{eq:losses_amplified_1}
\hat{\rho}_{\mathcal{A}}^{1 \psi}(T_{1}, T_{2}) &= \sum_{i=0}^{\infty} \sum_{j=0}^{i-1}
\sum_{k=i-j}^{\infty} \big[ \hat{\rho}_{\mathcal{A}}^{1 \psi}(T_{1}, T_{2}) 
\big]_{ijk}^{(i \geq j+1)} \vert i \psi \rangle_{1} \langle k \psi \vert \otimes
\vert j \psi_{\bot} \rangle_{2} \langle (j+k-i)\psi_{\bot}\vert +\\
&+ \sum_{i=0}^{\infty} \sum_{j=i}^{\infty} \sum_{k=0}^{\infty} \big[ 
\hat{\rho}_{\mathcal{A}}^{1 \psi}(T_{1}, T_{2}) \big]_{ijk}^{(i \geq j)} 
\vert i \psi \rangle_{1} \langle k \psi \vert \otimes \vert j \psi_{\bot} 
\rangle_{2} \langle (j+k-i)\psi_{\bot}\vert
\end{aligned}
\end{equation}
where the coefficients for $i\geq j+1$ and $i\leq j$ are given by:
\begin{eqnarray}
\big[ \hat{\rho}_{\mathcal{A}}^{1 \psi}(T_{1}, T_{2}) 
\big]_{ijk}^{(i \geq j)} &=& \frac{1}{C^{4}} \Gamma^{i+k-2}  \frac{T_{1}^{(i+k)/2}
T_{2}^{(2j+k-i)/2} R_{2}^{i-j-1} \sqrt{i!k!}}{(i-j-1)! \sqrt{j!(j+k-i)!}} 
\, _{2}F_{1}\left(1+i,1+k,i-j;\Gamma^{2} R_{1} R_{2} \right)\\
\label{eq:losses_amplified_3}
\big[ \hat{\rho}_{\mathcal{A}}^{1 \psi}(T_{1}, T_{2}) 
\big]_{ijk}^{(i < j)} &=& \frac{1}{C^{2}} \Gamma^{i+k} (j+1)(j+k-i+1) \frac{T_{1}^{(i+k)/2}
T_{2}^{(2j+k-i)/2} R_{1}^{j-i+1} \sqrt{j!(j+k-i)!}}{(j-i+1)! \sqrt{i!k!}} \\
&& \nonumber \, _{2}F_{1}\left(2+j,2+j+k-i,2+j-i;\Gamma^{2} R_{1} R_{2} \right)
\end{eqnarray}
\end{widetext}
According to previous considerations, the density matrices of
the states $\hat{\rho}_{\mathcal{A^{\prime}}}^{0} (T_{1},T_{2})$ and $\hat{%
\rho}_{\mathcal{A^{\prime}}}^{1 \psi_{\bot}} (T_{1},T_{2})$ for amplifier $%
\mathcal{A^{\prime}}$ can be directly derived from Eqs.(\ref%
{eq:losses_spontaneous_1}-\ref{eq:losses_spontaneous_3}) and (\ref%
{eq:losses_amplified_1}-\ref{eq:losses_amplified_3}) by substituting $(
\Gamma)$ with $(- \Gamma)$ and by re-labelling the indexes describing the
spatial and polarization modes.
Finally, the complete output state can be reconstructed as:
\begin{equation}
\hat{\rho}^{1\psi}(T_{1},T_{2}) = \hat{\rho}_{\mathcal{A}}^{1\psi}(T_{1},T_{2}) \otimes
\hat{\rho}_{\mathcal{A}'}^{0}(T_{1},T_{2})
\end{equation}

\subsection{Density matrix coefficients on the reduced $\mathbf{k}_{1}$ spatial mode
for the amplified states}

In this section we report the expression of the coefficients for the reduced
density matrix on spatial mode $\mathbf{k}_{1}$ of the $\vert \Phi^{1\psi} \rangle$ 
after the propagation over a lossy channel. Such result has been exploited in the
calculation of the Bures distance, where the action of the 
O-Filter device has been analyzed. The starting point of the calculation is the expression
(\ref{eq:lossy_channel_map}) of the $\vert \Phi^{1\psi} \rangle$. After the partial trace on 
mode $\mathbf{k}_{2}$, the density matrix $\hat{\rho}_{\mathbf{k}_{1}}^{1\psi} = 
\mathrm{Tr}_{\mathbf{k}_{2}} \left[Ê\vert \Phi^{1\psi} \rangle \langle \Phi^{1 \psi} \vert \right]$ reads:
\begin{equation}
\begin{aligned}
\hat{\rho}_{\mathbf{k}_{1}}^{1\psi} &= \sum_{n=0}^{\infty} \sum_{m=0}^{\infty} 
\frac{\Gamma^{2n+2m}}{C^{6}} (n+1) \\
&\vert (n+1)\psi 	\rangle_{1} \langle (n+1)\psi \vert \otimes
\vert m\psi_{\bot} \rangle_{1} \langle m\psi_{\bot} \vert 
\end{aligned}
\end{equation}
Finally, the application of the lossy channel map leads to the following density matrix:
\begin{equation}
\hat{\rho}^{1\psi}_{\mathbf{k_{1}}}(T) = \sum_{i=0}^{\infty}	\sum_{j=0}^{\infty}\left[ \rho^{1\psi}_{\mathbf{k}_{1}} (T) \right]_{ij} \vert i\psi \rangle_{1} \langle i\psi 
\vert \otimes \vert j\psi_{\bot} \rangle_{1} \langle j\psi_{\bot} \vert
\end{equation}
where the coefficients are given by:
\begin{equation}
\begin{aligned}
\left[ \rho_{\mathbf{k_{1}}}^{1\psi}\right]_{ij} &= \frac{\Gamma^{2i+2j-2}}{C^{6}} \eta^{i+j}
\left( i + \Gamma^{2} (1-\eta) \right) \\
& \left( 1 - \Gamma^{2} (1-\eta) \right)^{-3-i-j}
\end{aligned}
\end{equation}

\end{appendix}


\begin{thebibliography}{99}

\bibitem{Schr35}
E.~Schrodinger,
\newblock Naturwissenschaften {\bf 23}, 807 (1935).

\bibitem{Zure03}
W.~H. Zurek,
\newblock Rev. Mod. Phys. {\bf 75}, 715 (2003).

\bibitem{DeMa08}
F.~{De Martini}, F.~Sciarrino, and C.~Vitelli,
\newblock Phys. Rev. Lett. {\bf 100}, 253601 (2008).

\bibitem{Schl91}
W.~Schleich, M.~Pernigo, and F.~{Le Kien},
\newblock Phys. Rev. A {\bf 44}, 2172 (1991).

\bibitem{Gree89}
D.~M. Greenberger, M.~A. Horne, and A.~Zeilinger,
\newblock {\em Bell's theorem, Quantum Theory, and Conceptions of the Universe}
  (Kluwer Academic Publisher, 1989).

\bibitem{DeMa98}
F.~{De Martini},
\newblock Phys. Rev. Lett. {\bf 81}, 2842 (1998).

\bibitem{DeMa98a}
F.~{De Martini},
\newblock Phys. Lett. A {\bf 250}, 15 (1998).

\bibitem{Naga07}
E.~Nagali, T.~{De Angelis}, F.~Sciarrino, and F.~{De Martini},
\newblock Phys. Rev. A {\bf 76}, 042126 (2007).

\bibitem{DeMa09}
F.~{De Martini}, F.~Sciarrino, and N.~Spagnolo,
\newblock Phys. Rev. A {\bf 79}, 052305 (2009).

\bibitem{DeMa09a}
F.~{De Martini}, F.~Sciarrino, and N.~Spagnolo,
\newblock Phys. Rev. Lett. {\bf 103}, 100501 (2009).

\bibitem{DeMa05}
F.~{De Martini} and F.~Sciarrino,
\newblock Progr. Quantum Electron. {\bf 29}, 165 (2005).

\bibitem{DeMa05a}
F.~{De Martini}, F.~Sciarrino, and V.~Secondi,
\newblock Phys. Rev. Lett. {\bf 95}, 240401 (2005).

\bibitem{DeMa07}
F.~{De Martini} and F.~Sciarrino,
\newblock J. Phys. A: Math. Theor. {\bf 40}, 2977 (2007).

\bibitem{Bure69}
D.~Bures,
\newblock Trans. Am. Math. Soc. {\bf 135}, 199 (1969).

\bibitem{Hubn92}
M.~Hubner,
\newblock Phys. Lett. A {\bf 163}, 239 (1992).

\bibitem{Jozs94}
R.~Jozsa,
\newblock J. Mod. Opt. {\bf 41}, 2315 (1994).

\bibitem{Pell03}
D.~Pelliccia, V.~Schettini, F.~Sciarrino, C.~Sias, and F.~{De Martini},
\newblock Phys. Rev. A {\bf 68}, 042306 (2003).

\bibitem{DeMa04}
F.~{De Martini}, D.~Pelliccia, and F.~Sciarrino,
\newblock Phys. Rev. Lett. {\bf 92}, 067901 (2004).

\bibitem{DeMa02}
F.~{De Martini}, V.~Buzek, F.~Sciarrino, and C.~Sias,
\newblock Nature {\bf 419}, 815 (2002).

\bibitem{Pere95}
A.~Peres,
\newblock {\em Quantum Theory: Methods and Concepts} (Kluwer Academic
  Publisher, 1995).

\bibitem{Hutt96}
B.~Huttner, A.~Muller, J.~D. Gautier, H.~Zbinden, and N.~Gisin,
\newblock Phys. Rev. A {\bf 54}, 3783 (1996).

\bibitem{Scia05}
F.~Sciarrino and F.~{De Martini},
\newblock Phys. Rev. A {\bf 72}, 062313 (2005).

\bibitem{Scia07}
F.~Sciarrino and F.~{De Martini},
\newblock Phys. Rev. A {\bf 76}, 012330 (2007).

\bibitem{Cai08} J. Cai, S. Popescu, and H.-J. Briegel,  arXiv:0809.4906

\bibitem{Enge07} G.S. Engels, \emph{et al.}, Nature \textbf{446}, 782 (2007)

\bibitem{Durk04}
G.~A. Durkin, C.~Simon, J.~Eisert, and D.~Bouwmeester,
\newblock Phys. Rev. A {\bf 70}, 062305 (2004).

\bibitem{Slat66}
L.~J. Slater,
\newblock {\em Generalyzed Hypergeometric Functions} (Cambridge University
  Press, 1966). 

\end{thebibliography}
\end{document}